\documentclass[%
 aip,
 jmp,%
 amsmath,amssymb,
preprint,%
]{revtex4-1}

\usepackage{graphicx}% Include figure files
\usepackage{dcolumn}% Align table columns on decimal point
\usepackage{bm}% bold math
\def\be{\begin{equation}}
\def\ee{\end{equation}}
\def\bea{\begin{eqnarray}}
\def\eea{\end{eqnarray}}
\def\eq#1{(\ref{#1})}
\begin{document}

\preprint{YITP-11-45}

\title[ Tensor models and 3-ary algebras]{ Tensor models and 3-ary algebras}

\author{Naoki Sasakura}
\email{sasakura@yukawa.kyoto-u.ac.jp}
\affiliation{ Yukawa Institute for Theoretical Physics, Kyoto University,\\
Kyoto 606-8502, Japan
%\\This line break forced with \textbackslash\textbackslash
}%

\date{\today}% It is always \today, today,

\begin{abstract}
Tensor models are the generalization of matrix models, and are studied 
as models of quantum gravity in general dimensions.
In this paper, I discuss the algebraic structure in the 
fuzzy space interpretation of the tensor models which have a tensor with three indices
as its only dynamical variable.
The algebraic structure is studied mainly from the 
perspective of 3-ary algebras. It is shown that the tensor models have algebraic expressions, 
and that their symmetries are represented by 3-ary algebras. 
It is also shown that the 3-ary algebras of coordinates, which appear in
the nonassociative 
fuzzy flat spacetimes corresponding to a certain class of configurations with Gaussian functions in the tensor models,
form Lie triple systems, and the associated Lie algebras are shown to agree with those of  
the Snyder's noncommutative spacetimes.
The Poincare transformations of the coordinates
 on the fuzzy flat spacetimes are shown to be generated by 3-ary algebras.
\end{abstract}

%Use showkeys class option if keyword
\maketitle

\section{Introduction}
\label{sec:intro}
Tensor models have originally been introduced \cite{Ambjorn:1990ge,Sasakura:1990fs,Godfrey:1990dt}
to describe the simplicial quantum gravity in more
than two dimensions, with the hope to extend the successful description of the two-dimensional 
simplicial quantum gravity by the matrix models to other dimensions. 
The formulation has later been extended \cite{Boulatov:1992vp,Ooguri:1992eb,DePietri:1999bx}
to describe spin foam and loop quantum gravities by
considering Lie-group valued indices\footnote{This kind of models are also called group field theory.}. 
Despite various difficulties \cite{DePietri:2000ii}  
and the rather slow development since the introduction of tensor models, 
some interesting results have been reported recently 
\cite{Gurau:2011xq,Gurau:2010ba,Geloun:2010vj,Gurau:2010nd, Gurau:2009tz,Gurau:2009tw,Baratin:2011tg}.
These developments strengthen the general belief that
tensor models indicate the right direction to the background independent formulation of 
quantum gravity. 

Concerning the background independence of quantum gravity, the dimensions of spacetime
should be regarded as an effective dynamical quantity rather than a given constant.
This viewpoint not only seems natural from the physical requirement of quantum gravity, 
but has also been supported by some recent results from simplicial 
quantum gravity \cite{Ambjorn:2010rx} and field theoretical treatment \cite{Lauscher:2005qz}.
On the other hand, however, the original formulation of tensor models as well as the group field theory 
depend on the considering dimensions in their formalism through the rank of tensors and the choices of groups.
Therefore it would be desired to find another interpretation of tensor models 
which singles out a tensor model that is applicable to general dimensions.

The simplest choice of a tensor model is to consider one 
which has a tensor with three indices as its only dynamical variable. 
Then, by identifying the rank-three tensor with the structure constant of an algebra 
charactering a fuzzy space, the tensor model can be interpreted as 
theory of a dynamical fuzzy space.
Since one can in principle choose the values of the rank-three tensor to construct fuzzy spaces 
corresponding to any dimensional spaces, the rank-three tensor models can equally treat
spaces in general dimensions. 
This idea has first been presented in Ref.~\onlinecite{Sasakura:2005js}, 
and the subsequent studies mainly in numerical methods have supported the validity of this basic idea 
\cite{Sasakura:2005gv,Sasakura:2006pq,Sasakura:2007sv,Sasakura:2007ud,Sasakura:2008pe,Sasakura:2009hs,
Sasakura:2009dk,Sasakura:2010rb}.
The purpose of the present paper is to provide a full treatment of the 
original incomplete presentation of the idea, and 
to pursue the algebraic description of the tensor models.
In the sequel, it is found that 
3-ary algebras \cite{FigueroaO'Farrill:2008bd,deAzcarraga:2010mr,deAzcarraga:2011sh} 
describe the symmetries of the tensor models.
3-ary algebras have been introduced in physics by Nambu \cite{Nambu:1973qe}, and 
have recently been widely discussed in the context of M-theory 
\cite{Bagger:2006sk,Gustavsson:2007vu,Bagger:2007jr}.
This unexpected common appearance of 3-ary algebras suggests the general importance of 
this new way of describing symmetry in the physics of quantum spacetime.

This paper is organized as follows.
In the following section, the rank-three tensor model is presented.
In Section \ref{sec:algebra}, the structure of the algebras corresponding to 
the rank-three tensor models is discussed. 
In Section \ref{sec:commutative}, the commutative case of the algebras is discussed.
In Section \ref{sec:algebraicdescription}, the rank-three tensor model
is described in terms of the algebras.
In Section \ref{sec:algebragauss}, the fuzzy flat spacetimes with the algebras corresponding 
to a certain class of configurations with Gaussian functions in the rank-three tensor models
are discussed.    
In Section \ref{sec:snyder}, the 3-ary algebra of the coordinates obtained in the previous section
is studied, and a connection with the Snyder's noncommutative spacetime is found.
In Section \ref{sec:poincare}, the Poincare symmetry of the fuzzy flat spacetimes is shown to be 
represented by 3-ary operations.
The final section is devoted to summary, discussions and future prospects. 

\section{The rank-three tensor models}
\label{sec:tensormodel}
The simplest generalization of the matrix models will be the tensor models which have
a tensor with three indices $M_{abc}\ (a,b,c=1,2,\ldots,N)$ as their only dynamical variable
\cite{Ambjorn:1990ge,Sasakura:1990fs,Godfrey:1990dt}. 
The $M_{abc}$ takes complex values, and 
the hermiticity of a matrix  is generalized to 
\be
M_{abc}=M_{bca}=M_{cab}=M^*_{bac}=M^*_{acb}=M^*_{cba}, 
\label{eq:ghermiticity}
\ee
which states that the tensor takes its complex conjugate under  
odd permutations of the indices, while they are unchanged under even permutations.  

Because of the property \eq{eq:ghermiticity}, the symmetry which can be associated to the tensor models is 
the real orthogonal group symmetry,
\be
M_{abc}\rightarrow O_a{}^{a'} O_{b}{}^{b'} O_{c}{}^{c'} M_{a'b'c'},\ \ \  O\in O(N,R),
\label{eq:symtensor}
\ee
instead of the unitary groups for the hermitian matrix models.

Let me define a new tensor,
\be
C_{abc}=M_{abc}+M_{bac}+i (M_{abc}-M_{bac}).
\label{eq:cfromm}
\ee
Because of the generalized hermiticity condition \eq{eq:ghermiticity}, this tensor is real,
and is invariant under the cyclic permutations of the indices,
\be
C_{abc}=C_{bca}=C_{cab}.
\label{eq:ccyclic}
\ee
Note that $C_{abc}$ has no dependence under the odd permutations of the indices.
The symmetry transformation for $C_{abc}$ is given by a formula similar to  \eq{eq:symtensor}.

In fact, $C_{abc}$ is equivalent to $M_{abc}$ as degrees of freedom, because, 
from \eq{eq:cfromm}, one can show that $M_{abc}$ can also be expressed by $C_{abc}$ as 
\be
M_{abc}=\frac{1}{4}(C_{abc}+C_{bac})-\frac{i}{4}(C_{abc}-C_{bac}).
\ee
Therefore, a tensor model with real $C_{abc}$ satisfying \eq{eq:ccyclic} 
is equivalent to the
one with complex $M_{abc}$ satisfying \eq{eq:ghermiticity}.
In the following discussions, real $C_{abc}$ with \eq{eq:ccyclic} will be used.

\section{Properties of corresponding algebras}
\label{sec:algebra}
In this section, I will discuss an interpretation of the tensor models in terms of
the notion of fuzzy spaces.

A fuzzy space is characterized by an algebra of functions $\phi_a\,(a=1,2,\ldots,N)$ on it, which form
a basis.  Generally, the multiplication of such an algebra is defined by a structure constant $f_{ab}{}^c$
as 
\be
\phi_a \phi_b=f_{ab}{}^c \phi_c,
\label{eq:ppfp}
\ee
where $f_{ab}{}^c$ is assumed to be real to describe a real fuzzy space.
It is important to 
note that the algebra may not be associative; rather,
nonassociativity will play essential roles in the discussions of this paper\footnote{In this paper, an algebra is allowed to be nonassociative in general. 
This is physically required to circumvent the Wedderburn's theorem, which roughly requires a finite associative
algebra
to be a matrix algebra. A matrix algebra would be too simple to represent a fuzzy space with geometric 
properties.}.  

I also assume there exists a metric, 
\be
\langle \phi_a|\phi_b\rangle=h_{ab},
\ee
which is assumed to be real, symmetric, $h_{ab}=h_{ba}$, and bilinear.

The basis of functions $\phi_a$ can be changed to another one. 
A new basis may be given by functions of $\phi_a$. By using \eq{eq:ppfp},
any nonlinear function of $\phi_a$ can be expressed linearly, because 
\bea
\phi_a'&=&A_{a}{}^b \phi_b+A_{a}{}^{bc}\phi_b\phi_c+\cdots \nonumber \\
&=& (A_a{}^d+A_a{}^{bc}f_{bc}{}^d+\cdots)\phi_d,
\eea
where $A$'s are real numerical coefficients.
Therefore, any new basis provided by functions of $\phi_a$ 
can in fact be obtained by a linear transformation of $\phi_a$.
Since the change of basis should be invertible, 
it is given by a general linear group transformation,
\be
\phi_a\rightarrow M_a{}^{a'}\phi_{a'},\ \ \ M\in GL(N,R).
\label{eq:symalg}
\ee
Under this general linear group transformation, the $f_{ab}{}^c$ and $h_{ab}$ are 
transformed by
\bea
f_{ab}{}^c&\rightarrow& M_a{}^{a'}M_b{}^{b'}M^{-1}_{c'}{}^c f_{a'b'}{}^{c'},\nonumber \\
h_{ab}&\rightarrow& M_a{}^{a'}M_b{}^{b'}h_{a'b'}.
\label{eq:transfh}
\eea
The above transformations of the basis functions may be interpreted as the fuzzy space analogue to 
the diffeomorphisms of an ordinary space, since a diffeomorphism generates a linear transformation 
of a basis of functions on it \cite{Sasakura:2005js}.

To relate the above algebraic structure with the rank-three
tensor models, let me assume a relation,
\be
C_{abc}=f_{ab}{}^{c'}h_{c'c}.
\label{eq:cfrel}
\ee
Then the property of the cyclic symmetry \eq{eq:ccyclic} can be translated to the following 
cyclic condition on the algebra as
\be
\langle \phi_a \phi_b |\phi_c \rangle=\langle \phi_a | \phi_b \phi_c\rangle=\langle \phi_b | \phi_c \phi_a \rangle.
\label{eq:cyclicalg}
\ee
The former equality is because 
\bea
\langle \phi_a \phi_b |\phi_c \rangle&=&f_{ab}{}^d \langle \phi_d | \phi_c \rangle=f_{ab}{}^d h_{dc}=C_{abc}, \nonumber \\
\langle \phi_a | \phi_b \phi_c\rangle&=& f_{bc}{}^d \langle \phi_a | \phi_d \rangle =f_{bc}{}^d h_{ad} =C_{bca}.
\eea
The proof of the latter is similar.

Now let me define a 3-ary product 
by\footnote{Because of the possible nonassociativity of the algebra, the 
order of the products must be explicitly indicated.}
\be
[\phi_a,\phi_b;\phi_c]\equiv(\phi_a \phi_c)\phi_b-(\phi_b \phi_c)\phi_a.
\label{eq:3ary}
\ee
The first two entries are antisymmetric,
\be
[\phi_a,\phi_b;\phi_c]=-[\phi_b,\phi_a;\phi_c].
\label{eq:3aryanti}
\ee
The metric is invariant under the 3-ary operation on the third entry,
\be
\langle [\phi_a,\phi_b;\phi_c]|\phi_d\rangle=-\langle \phi_c | [\phi_a,\phi_b;\phi_d]\rangle.
\label{eq:metricinv}
\ee
This can be proved by using \eq{eq:cyclicalg} as
\bea
\langle [\phi_a,\phi_b;\phi_c]|\phi_d\rangle&=&\langle (\phi_a\phi_c)\phi_b-(\phi_b\phi_c)\phi_a|\phi_d\rangle 
\nonumber \\
&=&\langle (\phi_a\phi_c)\phi_b|\phi_d\rangle-\langle(\phi_b\phi_c)\phi_a|\phi_d\rangle 
\nonumber \\
&=& \langle \phi_a \phi_c |\phi_b \phi_d\rangle - \langle \phi_b \phi_c| \phi_a \phi_d\rangle \nonumber \\
&=& \langle \phi_c|(\phi_b\phi_d)\phi_a\rangle - \langle \phi_c | (\phi_a\phi_d)\phi_b\rangle \nonumber \\
&=& -\langle \phi_c |[\phi_a,\phi_b;\phi_d]\rangle.
\eea 

This invariance can be discussed more explicitly by writing down the linear transformation
associated to \eq{eq:3ary} as
\bea
[\phi_a,\phi_b;\phi_c]
&=&(\phi_a \phi_c)\phi_b-(\phi_b \phi_c)\phi_a \nonumber \\
&=& f_{ac}{}^d f_{db}{}^e\phi_e -f_{bc}{}^d f_{da}{}^e \phi_e \nonumber \\
&=& (M_{ab})_c{}^d \phi_d, 
\label{eq:3arytrans}
\eea
where
\be
(M_{ab})_c{}^{d}=f_{ac}{}^e f_{eb}{}^d -f_{bc}{}^e f_{ea}{}^d.
\ee
The indices of $(M_{ab})_c{}^d$ have the following anti-symmetric properties,  
\bea
(M_{ab})_c{}^d&=&-(M_{ba})_c{}^d, \nonumber \\
(M_{ab})_{cd}&=&-(M_{ab})_{dc},
\label{eq:propertyofm}
\eea
where $(M_{ab})_{cd}=(M_{ab})_c{}^{d'} h_{d'd}$. 
The latter equation can be proven by the relation \eq{eq:cfrel} and the cyclic property \eq{eq:ccyclic}.

Especially, the latter equation of \eq{eq:propertyofm}
shows that $h_{ab}$ is invariant under the infinitesimal transformation
\eq{eq:3arytrans}. Therefore there exist various invariant quantities under the transformation. 
For example, if $h_{ab}$ is invertible, 
\be
h^{ab}\phi_a \phi_b
\label{eq:inv1}
\ee
is invariant, where $h^{ab}$ is the inverse of $h_{ab}$. 
It is clear that 
one can construct various invariants by contracting the lower indices of $\phi_a$ with $h^{ab}$.

Another kind of invariant is the trace defined by
\be
\hbox{Tr}({\cal O})\equiv h^{ab}\langle \phi_a | {\cal O} \phi_b\rangle=h^{ab} \langle \phi_a {
\cal O}|\phi_b\rangle.
\ee
When ${\cal O}$ is an invariant, the whole expression ${\rm Tr}( {\cal O})$ is also invariant.
The latter equation is valid, because of \eq{eq:cyclicalg}.

It should be stressed that, due to the general possibility of nonassociativity of the algebra, 
the 3-ary operation \eq{eq:3arytrans} on the third entry does not satisfy the Leibnitz rule in general\footnote{However, for example, 
see Ref.~\onlinecite{Axenides:2008rn} and the references therein for 
some efforts to realize the Leibnitz rule in quantization of Nambu bracket.}.
Therefore, for example, the invariant \eq{eq:inv1} is not invariant under the 3-ary operation
to the whole expression,
\be
[\phi_a,\phi_b;h^{cd}\phi_c \phi_d]\neq 0\ \ \hbox{in general},
\ee
while the following equation, 
\be
h^{cd} [\phi_a,\phi_b;\phi_c] \phi_d+h^{cd} \phi_c [\phi_a,\phi_b;\phi_d]=0,
\ee 
holds.

The orthogonal group symmetry \eq{eq:symtensor} of the tensor models and the general
linear symmetry \eq{eq:symalg}
of the algebra can be identified by partial gauge fixing of the latter symmetry as follows.
Let me assume that the metric $h_{ab}$ is positive definite. Then the transformation
\eq{eq:transfh} allows $h_{ab}$ to be gauge fixed to 
\be
h_{ab}=\delta_{ab}.
\label{eq:hdelta}
\ee
Under the gauge fixing \eq{eq:hdelta}, the remaining symmetry agrees with 
the orthogonal group symmetry \eq{eq:symtensor} of the tensor models.
In this case, the Lie group generated by the infinitesimal transformation \eq{eq:3arytrans}
is a subgroup of the orthogonal group $O(N,R)$ in general. 

With the gauge fixing \eq{eq:hdelta}, 
the structure constant of the algebra and the dynamical variable of the tensor models 
can be identified by
\be
C_{abc}=f_{ab}{}^c.
\label{eq:ceqf}
\ee
Thus, the degrees of freedom of fuzzy spaces and those of the tensor models coincide.

\section{Imposing commutativity of the algebra}
\label{sec:commutative}
In the following discussions, let me assume that the tensor $C_{abc}$ is not only cyclic symmetric but 
totally symmetric,
\be
C_{abc}=C_{bca}=C_{cab}=C_{bac}=C_{acb}=C_{cba}.
\label{eq:csymcond}
\ee
This reduction of the degrees of freedom has been introduced previously \cite{Sasakura:2006pq} 
to simplify the analysis of the tensor models. In fact, this reduction of the degrees of freedom 
does not diminish the physical interests in the tensor models, since 
various interesting properties in relation with the general relativity have been shown 
\cite{Sasakura:2007sv,Sasakura:2007ud,Sasakura:2008pe,Sasakura:2009hs,
Sasakura:2009dk,Sasakura:2010rb} under the reduction.

In addition to the property \eq{eq:cyclicalg}, the total symmetry requires that the algebra be commutative,
\be
\phi_a \phi_b=\phi_b \phi_a.
\label{eq:commutativity}
\ee

While the commutators between $\phi_a$ vanish on account of the commutativity \eq{eq:commutativity},
the 3-ary product \eq{eq:3ary} is equivalent to an associator \cite{Schafer},
\be
[\phi_a,\phi_b;\phi_c]=(\phi_a \phi_c)\phi_b-\phi_a(\phi_c \phi_b),
\label{eq:associator}
\ee
and takes non-vanishing values in general, reflecting the nonassociativity of the algebra.

In the commutative case, the following cyclic identity holds,
\be
[\phi_a,\phi_b;\phi_c]+[\phi_b,\phi_c;\phi_a]+[\phi_c,\phi_a;\phi_b]=0.
\label{eq:jacobiid}
\ee
This can be shown from \eq{eq:commutativity} and \eq{eq:associator}.

As in the general non-commutative case discussed in the previous section, 
the associator \eq{eq:associator} does not satisfy the Leibnitz rule in general also for the commutative case.
However, there exists a physically interesting 3-Leibnitz subalgebra \cite{deAzcarraga:2010mr,deAzcarraga:2011sh} 
which satisfies a kind of Leibnitz rule and 
is intimately related to the tensor models. This will be discussed in Section \ref{sec:snyder} .  

\section{Algebraic description of the rank-three tensor models}
\label{sec:algebraicdescription}
After the gauge fixing \eq{eq:hdelta}, the structure constant of the algebra
and the dynamical variable of the tensor models can be identified as in \eq{eq:ceqf}.
Therefore, it is possible to write down actions of the tensor models in terms of the algebra.

For simplicity, let me assume the total symmetry \eq{eq:csymcond} of $C_{abc}$, which requires
the commutativity \eq{eq:commutativity} of the algebra. 
It should be straightforward to extend the following discussions also to the general cases
with noncommutativity. 
In the quadratic order of $C_{abc}$, there exist two actions which are invariant under the orthogonal group 
symmetry \eq{eq:symtensor}, and they can be expressed with the algebraic language as\footnote{The repeated indices
are assumed to be summed over.}
\bea
S^{(2)}_1 &=&C_{abc} C_{abc}=\langle \phi_a \phi_b |\phi_a \phi_b \rangle, \nonumber \\
S^{(2)}_2&=&C_{aac}C_{bbc}=\langle \phi_a \phi_a | \phi_b \phi_b \rangle.
\eea
These equations can be checked by using $\phi_a \phi_b=C_{abc}\phi_c$ under the identification \eq{eq:ceqf}.

More interesting forms appear in the quartic order. For example,
\bea
S_1^{(4)}&=&C_{abc}C_{abd}C_{efc}C_{efd}=\langle (\phi_a \phi_b) \phi_c | (\phi_a \phi_b) \phi_c\rangle,\nonumber \\
S_2^{(4)}&=&C_{abc}C_{ade}C_{bdf}C_{cef}=\langle (\phi_a \phi_b) \phi_c | \phi_a (\phi_b \phi_c)\rangle.
\label{eq:s4}
\eea

A usage of this kind of algebraic expression is to find an action bounded from below to 
assure the stability of the tensor models.
The action $S_1^{(4)}$ is positive definite, because it has a form of a norm square for the positive definite 
metric \eq{eq:hdelta}. 
On the other hand, $S_2^{(4)}$ is not so, but can be made it bounded from below by combining with $S_1^{(4)}$,
as can be proven by 
\be
S_1^{(4)}+S_2^{(4)}=\frac12\langle  (\phi_a \phi_b) \phi_c + \phi_a (\phi_b \phi_c) |
  (\phi_a \phi_b) \phi_c + \phi_a (\phi_b \phi_c) \rangle
  \geq 0.
\ee
It is obvious that one can construct various actions by considering such invariants of the algebra. 

As described in Section \ref{sec:tensormodel}, the symmetry of such an action of a tensor model 
is given by the orthogonal group symmetry $O(N,R)$. On the other hand, 
as discussed in Section \ref{sec:algebra}, 
the 3-ary product \eq{eq:3ary} can be regarded as the generators of this symmetry under the 
gauge fixing condition \eq{eq:hdelta} and the identification \eq{eq:ceqf}.
Therefore the symmetry of the tensor models can be     
written algebraically by using the 3-ary transformation 
\eq{eq:3arytrans} as
\be
\delta_{ab} \phi_c=[\phi_a,\phi_b;\phi_c].
\label{eq:tensortransalg}
\ee
Since, on account of \eq{eq:3aryanti}, 
the number of the independent choices of $\phi_a$ and $\phi_b$ agrees with the dimension of the orthogonal group
$O(N,R)$, the 3-ary transformations \eq{eq:tensortransalg} will span all the symmetry generators unless
the configuration $C_{abc}$ is fine-tuned not to be 
so\footnote{More detailed discussions will be given elsewhere \cite{Sasakura:2011nj}.}. 
In fact, by choosing appropriate elements $\phi_a$ and $\phi_b$, various transformations of physical interests 
can be constructed.
In the following sections, Poincare transformations of the coordinates 
of fuzzy flat spacetimes will explicitly be given.  

\section{The algebra corresponding to the Gaussian configurations in the tensor models}
\label{sec:algebragauss}
In the study of emergent general relativity from the tensor models, a certain kind of configurations
in the tensor models play important roles \cite{Sasakura:2007sv,Sasakura:2007ud,Sasakura:2008pe,Sasakura:2009hs,
Sasakura:2009dk,Sasakura:2010rb}. 
These configurations have Gaussian forms given by
\be
C_{p^1\, p^2\, p^3}=\exp\left(-\alpha \left((p^1)^2+(p^2)^2+(p^3)^2\right)\right) \delta^D(p^1+p^2+p^3),
\label{eq:gaussconf}
\ee
where the indices are $D$-dimensional momentum, $p^i=(p^i_1,p^i_2,\ldots,p^i_D)$, and
$\delta^D(\cdot)$ denotes the $D$-dimensional $\delta$-function. 
Here $(p)^2$ denotes the square of $D$-dimensional momentum defined by
\be
(p)^2=g^{\mu\nu}p_\mu p_\nu,
\ee
where $g^{\mu\nu}$ is a constant real symmetric two-tensor.
The parameter $\alpha$ is redundant in the sense that it can be absorbed into the redefinition of $g^{\mu\nu}$, but it
is kept there for the dimensional reason.
These Gaussian configurations satisfy the totally symmetric condition
\eq{eq:csymcond} in the previous section, and therefore define the class of fuzzy spaces discussed so far.

From the identification \eq{eq:ceqf}, 
the algebra corresponding to \eq{eq:gaussconf} is given by\footnote{The metric in the coordinate basis is given by $h_{x_1x_2}=
\delta^D(x_1-x_2)$, which has the form of the gauge-fixing condition \eq{eq:hdelta}.
Then, by taking Fourier transformation of $h_{x_1x_2}$, 
the metric in the plane wave basis is given by $h_{p^1p^2}=\delta^D(p^1+p^2)$.},
\be
\phi_{p^1}\phi_{p^2}=C_{p^1p^2p^3}h^{p^3p^4} \phi_{p^4}
=\exp\left(-2 \alpha \left((p^1)^2+(p^2)^2+p^1\cdot p^2\right)\right) 
\phi_{p^1+p^2},
\label{eq:gaussalg}
\ee 
where $p^1\cdot p^2$ denotes the inner product, $g^{\mu\nu} p^1_\mu p^2_\nu$.
Because of the exponential factor in \eq{eq:gaussalg}, the algebra \eq{eq:gaussalg} may be 
regarded as a deformation of the algebra of plane waves, $\phi_p\sim e^{ipx}$, in an ordinary spacetime, 
$\phi_{p^1}\phi_{p^2}=\phi_{p^1+p^2}$.
The algebra \eq{eq:gaussalg} is commutative but nonassociative, and the parameter $\alpha$ characterizes the 
scale of the nonassociativity.
The algebra is obviously invariant under the Poincare group in $D$-dimensions, and therefore it is 
physically interpreted as an algebra defining 
a $D$-dimensional nonassociative fuzzy flat spacetime \cite{Sasai:2006ua}.

Now let me define the spacetime coordinates of this fuzzy spacetime. 
For the ordinary spacetime, the coordinates can be obtained by taking derivatives
of the plane waves with respect to the momentum and putting the momentum to zero, 
$x^\mu=\left.-i\frac{\partial }{\partial p_\mu}e^{ipx}\right|_{p=0}$. 
Following this ordinary manner, let me define coordinates by
\be
\hat x^\mu=-i\left.\frac{\partial \phi_p}{\partial p_\mu}\right|_{p=0} 
= i \int d^Dp\, \delta^\mu(p) \phi_p,
\label{eq:defx}
\ee
where
\be
\delta^\mu(p)\equiv\delta(p_1)\delta(p_2)\cdots \delta'(p_\mu)\cdots \delta(p_D).
\label{eq:deltamu}
\ee
Here $\delta'(p)$ denotes the first derivative of the delta function defined by
\be
\int dp\, \delta'(p) \phi_p=-\left.\frac{d\phi_p}{dp}\right|_{p=0}.
\ee

Similarly, for later convenience, let me generalize this to
\be
\hat x^{\mu_1 \mu_2 \ldots \mu_n}\equiv
\left. 
(-i)^n \frac{\partial^n \phi_p}{\partial p_{\mu_1} \partial p_{\mu_2} \cdots \partial p_{\mu_n}}
\right |_{p=0} 
= i^n \int d^Dp\, \delta^{\mu_1\mu_2\ldots \mu_n}(p) \phi_p,
\ee
where $\delta^{\mu_1\mu_2\ldots \mu_n}(p)$ is defined similarly as \eq{eq:deltamu}. 
Then, using \eq{eq:gaussalg},
the products of $\hat x^{\mu_1 \mu_2 \ldots \mu_n}$ are given by
\bea
\hat x^{\mu_1\mu_2\ldots \mu_n} \hat x^{\nu_1 \nu_2 \ldots \nu_m}
=i^{n+m} \int d^D q d^Dq \, \delta^{\mu_1\mu_2\ldots \mu_n}(p) \delta^{\nu_1\nu_2\ldots\nu_m}(q)
\, e^{-2 \alpha(p^2+q^2+p\cdot q)}\, \phi_{p+q}.
\label{eq:prodxx}
\eea
From \eq{eq:prodxx}, it is easy to explicitly derive some low order products as
\bea
\label{eq:p0x}
\phi_0 \hat x^\mu&=& \hat x^\mu, \\
\label{eq:pxx}
\hat x^\mu \hat x^\nu&=& 2 \alpha g^{\mu\nu} \phi_0 + \hat x^{\mu\nu}, \\
\label{eq:px2x}
\hat x^{\mu\nu} \hat x^\rho&=& 4 \alpha g^{\mu\nu} \hat x^\rho + 2 \alpha g^{\mu\rho} \hat x^\nu
+2 \alpha g^{\nu\rho} \hat x^\mu,
\eea
where 
\be
\phi_0=\phi_p|_{p=0}.
\label{eq:phi0}
\ee
Especially, from \eq{eq:p0x}, \eq{eq:pxx} and \eq{eq:px2x}, one obtains
\bea
[\hat x^\mu,\hat x^\nu;\hat x^\rho]&=&(\hat x^\mu \hat x^\rho) \hat x^\nu-(\hat x^\nu \hat x^\rho) \hat x^\mu 
\nonumber\\
&=& (2 \alpha g^{\mu\rho} \phi_0 +\hat x^{\mu\rho})\hat x^\nu-(2 \alpha g^{\nu\rho} \phi_0+\hat x^{\nu\rho}) 
\hat x^{\mu} \nonumber \\
&=&
4 \alpha (g^{\mu\rho}\,\hat x^\nu-g^{\nu\rho}\,\hat x^\mu).
\label{eq:3aryx}
\eea

\section{The 3-ary product of coordinates and Snyder's noncommutative spacetime}
\label{sec:snyder}
The 3-ary product defined in \eq{eq:3ary} does not satisfy the Leibnitz rule in general, but
there exist the possibilities that subalgebras satisfy kinds of Leibnitz rules. In fact,
one can explicitly check that the 3-ary product \eq{eq:3aryx} of the coordinates satisfies 
the fundamental identity \cite{Takhtajan:1993vr} 
(or Filippov identity \cite{deAzcarraga:2010mr,deAzcarraga:2011sh}), 
which is given by
\be
[\hat x^\mu,\hat x^\nu;[\hat x^\rho,\hat x^\delta;\hat x^\epsilon]]=[[\hat x^\mu,\hat x^\nu;\hat x^\rho],
\hat x^\delta;\hat x^\epsilon]
+[\hat x^\rho,[\hat x^\mu,\hat x^\nu;\hat x^\delta];\hat x^\epsilon]+[\hat x^\rho,\hat x^\delta;
[\hat x^\mu,\hat x^\nu;\hat x^\epsilon]].
\label{eq:xleibnitz}
\ee 
This kind of algebra is called a 3-Leibnitz algebra
\cite{deAzcarraga:2010mr,deAzcarraga:2011sh}.

The antisymmetry of the first two entries \eq{eq:3aryanti}, the cyclic identity \eq{eq:jacobiid},
and the fundamental identity \eq{eq:xleibnitz} shows that the coordinates $\hat x^\mu$ and the 3-ary
product \eq{eq:3ary} form a Lie triple system 
\cite{Okubo:2003uh,FigueroaO'Farrill:2008bd,deAzcarraga:2010mr,deAzcarraga:2011sh}. 
A Lie triple system is known to have an associated Lie algebra,
and the same Lie triple system as the present case is explained in detail as an example in 
Ref.~\onlinecite{Okubo:2003uh} 
(see also Refs.~\onlinecite{FigueroaO'Farrill:2008bd,deAzcarraga:2010mr,deAzcarraga:2011sh}).
The construction of a Lie algebra from the Lie triple system starts with formally defining 
some anti-symmetric commutators and new generators $\hat M^{\mu\nu}$ by relations,
\bea 
\label{eq:defM}
4\alpha \hat M^{\mu\nu}&\equiv& [\hat x^\mu,\hat x^\nu]=-[\hat x^\nu,\hat x^\mu],\\
\ [\hat M^{\mu\nu},\hat x^\rho]&=& 
-[\hat x^\rho,\hat M^{\mu\nu}]\equiv\frac{1}{4 \alpha} [\hat x^\mu,\hat x^\nu;\hat x^\rho]=g^{\mu\rho} 
\hat x^\nu- g^{\nu\rho}\hat x^\mu,
\label{eq:defMop}
\eea
where the last equation is from \eq{eq:3aryx}.
Here it is important to note that the commutator $[\ ,\ ]$ is only formally defined, and is not
defined by products $[a,b]=ab-ba$, which identically vanishes on account of the commutativity of the algebra. 
The consistency of the formal definitions of the commutators as a Lie algebra is guaranteed by
the antisymmetry of the first two entries \eq{eq:3aryanti}, the cyclic identity \eq{eq:jacobiid},
and the fundamental identity \eq{eq:xleibnitz}. 
Especially, from consistency, one obtaines
\be
[\hat M^{\mu\nu},\hat M^{\rho\sigma}]=g^{\mu\rho} \hat M^{\nu\sigma}- g^{\mu\sigma} \hat M^{\nu\rho}
- g^{\nu\rho}\hat M^{\mu\sigma} +  g^{\nu\sigma} \hat M^{\mu\rho}.
\label{eq:comM}
\ee
Therefore $\hat M^{\mu\nu}$ can be interpreted as the generators of the rotational (Lorentz) transformation of 
the coordinates $\hat x^\mu$ of a flat space(time) with an inverse metric $g^{\mu\nu}$.

The associated Lie algebra  defined by \eq{eq:defM}, \eq{eq:defMop} and \eq{eq:comM} 
agrees with that of the Snyder's noncommutative spacetime \cite{Snyder:1946qz}
\footnote{The parameter $4\alpha$ corresponds to $a^2$ in Ref.~\onlinecite{Snyder:1946qz}.}.
Derivation of the Snyder's noncommutative spacetime from a Lie triple system similarly as above 
has first been discussed in another kind of nonassociative spacetimes in 
Refs.~\onlinecite{Girelli:2010zw,Girelli:2010wi}.
The two kinds of nonassociative spacetimes have different structures, but derive the same algebra of coordinates.
This is probably because the coordinates defined in \eq{eq:defx} are evaluated around $p=0$,
and both of the nonassociative spacetimes have common infrared structures.
These nonassociative spacetimes show interesting transmutation from nonassociativity to noncommutativity:
the products of two $\hat x^\mu$ are commutative, but products of more than two coordinates 
show noncommutativity as in the case of the 3-ary product. 
Thus, 3-ary products and Lie triple systems open a new way of interpreting noncommutative spacetimes. 
As has been stressed in Refs.~\onlinecite{Girelli:2010zw,Girelli:2010wi}, this way of realizing noncommutative 
spacetimes can serve as a new solution to the problem of unwanted dimensions in noncommutative spacetimes. 
In Snyder's noncommutative spacetime, the algebra of the coordinates $\hat x^\mu$ does not close by themselves
and  $\hat M^{\mu\nu}$ appear as other coordinates.
Therefore one has to confront a rather hard problem of making $\hat M^{\mu\nu}$ physically invisible.
On the other hand, however, the $\hat M^{\mu\nu}$ introduced in \eq{eq:defM} is just a label for representing 
a 3-ary operation, and
does not represent any coordinates.

\section{Extension to Poincare symmetry}
\label{sec:poincare}
In the previous section, the rotational (Lorentz) transformation of the space(time) coordinates $\hat x^\mu$ 
has been shown to be realized by a 3-ary operation \eq{eq:defMop}. 
In this section, it will be shown that  the translational transformation of the coordinates of the 
fuzzy flat spacetime can also be represented by a 3-ary operation.

What appears to be very strange in discussing the translational transformation of the spacetime 
coordinates $\hat x^\mu$
is that the 3-ary algebra \eq{eq:3aryx} of the coordinates $\hat x^\mu$ does not apparently seem  
invariant under  
the naive translation, $\hat x^\mu\rightarrow \hat x^\mu+v^\mu$, where $v^\mu$ is a $c$-number vector.
In fact, the left-hand side of \eq{eq:3aryx} does not change under the naive translation, 
while the last expression of 
\eq{eq:3aryx} is shifted.
On the other hand, however, the plane wave algebra \eq{eq:gaussalg} 
is obviously invariant under the phase rotations,
\be
\phi_p\rightarrow e^{i\,p\cdot v} \phi_p.
\label{eq:transplane}
\ee 
This is actually the same as the phase rotation generated by the translations in an ordinary spacetime.
Therefore, to find the correct translational transformation
on the fuzzy flat spacetime,
one has to take good care of \eq{eq:transplane}. 

By the replacement \eq{eq:transplane}, the coordinates defined in \eq{eq:defx} will be transformed to
\be
\hat x^\mu\rightarrow -i\left.\frac{\partial e^{i\,p\cdot v}\phi_p}{\partial p_\mu}\right|_{p=0}
=\hat x^\mu+v^\mu \phi_0,
\label{eq:replacex}
\ee
where $\phi_0$ is defined in \eq{eq:phi0}.
Therefore the shift of the coordinates under the translation is actually 
given by $\delta \hat x^\mu=v^\mu \phi_0$, and not by a simple $c$-number
vector.
In fact, $\phi_0$ is not a trivial element, because, from \eq{eq:gaussalg},
\be
\phi_0 \phi_p=e^{-2 \alpha (p)^2} \phi_p.
\label{eq:0p}
\ee  
From \eq{eq:0p} and \eq{eq:prodxx}, one also has
\bea
\phi_0\phi_0&=&\phi_0,
\label{eq:phi00} \\
\label{eq:p0xx}
\phi_0\, \hat x^{\mu\nu}&=&4 \alpha g^{\mu\nu} \phi_0 + \hat x^{\mu\nu}.
\eea
Then the shift of the left-hand side of \eq{eq:3aryx} by \eq{eq:replacex}
can be computed as 
\bea
\delta [\hat x^\mu,\hat x^\nu;\hat x^\rho]&=&\delta \left( (\hat x^\mu \hat x^\rho)\hat x^\nu-
(\hat x^\nu \hat x^\rho)\hat x^\mu\right)\nonumber \\
&=& (v^\mu \phi_0 \hat x^\rho)\hat x^\nu
+(\hat x^\mu v^\rho \phi_0)\hat x^\nu+
(\hat x^\mu \hat x^\rho)v^\nu \phi_0\nonumber \\
&&\ \ \ \ \ -
(v^\nu \phi_0 \hat x^\rho)\hat x^\mu-
(\hat x^\nu v^\rho \phi_0)\hat x^\mu-
(\hat x^\nu \hat x^\rho)v^\mu \phi_0 \nonumber \\
&=&4 \alpha  g^{\mu\rho} v^\nu\phi_0-4 \alpha g^{\nu\rho}v^\mu\phi_0,
\eea
where I have used \eq{eq:p0x}, \eq{eq:pxx}, \eq{eq:phi00} and \eq{eq:p0xx}.
The last expression indeed agrees with the shift of the last expression of \eq{eq:3aryx}, and 
therefore \eq{eq:3aryx} is invariant under the translational symmetry \eq{eq:replacex} on
the fuzzy flat spacetime.

Thus, to properly incorporate the translational symmetry of the fuzzy flat spacetime,
the 3-ary algebra of the coordinates $\hat x^\mu$ should be extended to include 
the element $\phi_0$. 
From \eq{eq:p0x}, \eq{eq:pxx}, \eq{eq:phi00} and \eq{eq:p0xx},
one can explicitly obtain 
\bea
[\hat x^\mu,\phi_0;\hat x^\nu]=-[\phi_0,\hat x^\mu;\hat x^\nu]&=&4 \alpha g^{\mu\nu}\phi_0, \nonumber \\
\hbox{[The others with $\phi_0$]}&=&0,
\label{eq:xx0}
\eea
and the extended 3-ary algebra is closed.
Then one finds that the translational transformation \eq{eq:replacex} can be represented by
\be
\delta \hat x^\mu=[v_\nu \hat x^\nu,\phi_0;\hat x^\mu].
\ee

In fact, the 3-ary algebra extended with $\phi_0$ can be regarded as the 3-ary algebra of the kind \eq{eq:3aryx} 
extended with a new coordinate $\phi_0$ with a degenerate metric,
\be
g^{\mu0}=g^{0\mu}=g^{00}=0.
\label{eq:singmetric}
\ee 
This can be checked by comparing \eq{eq:3aryx} with \eq{eq:xx0}.
Therefore, the extended algebra also forms a Lie triple system.
Following the same procedure as in the previous section, the associated Lie algebra can be shown to contain
the Poincare Lie algebra,
\bea
\ [\hat T^\mu,\hat T^\nu]&=&0, \nonumber \\
\ [\hat M^{\mu\nu},\hat T^\rho]&=&g^{\mu\rho}\hat T^\nu-g^{\nu\rho}\hat T^\mu, \nonumber \\
\ [\hat M^{\mu\nu},\hat M^{\rho\sigma}]&=&g^{\mu\rho} \hat M^{\nu\sigma}- g^{\mu\sigma} \hat M^{\nu\rho}
- g^{\nu\rho}\hat M^{\mu\sigma} +  g^{\nu\sigma} \hat M^{\mu\rho},
\eea
where 
\bea
4 \alpha \hat T^\mu&\equiv&[\hat x^\mu, \phi_0]=-[\phi_0,\hat x^\mu] , \nonumber \\
\ [\hat T^\mu,\hat x^\nu]&=& -[\hat x^\nu,\hat T^\mu]\equiv 
\frac{1}{4 \alpha} [\hat x^\mu,\phi_0;\hat x^\nu]= g^{\mu\nu}\phi_0.
\eea

\section{Summary, discussions and future prospects}
In this paper, I have studied the idea, first presented in Ref.~\onlinecite{Sasakura:2005js}, 
that the rank-three tensor models can be interpreted as theory of dynamical fuzzy spaces.
The algebra of functions on a corresponding fuzzy space is shown to be constrained to have a certain cyclic property, 
which comes from the generalized hermiticity condition on the three-index tensor in the rank-three tensor models.  
The actions of the rank-three tensor models can be represented algebraically, and 
such algebraic description will provide a new tool to study the rank-three tensor models.

The most important implication of wide interest of this paper would be that the cyclic property of algebras
may be used to define an interesting class of fuzzy spaces. 
As discussed in Section \ref{sec:algebra}, the cyclic property is the essential feature to 
guarantee the 3-ary operation to be a transformation which conserves the metric of the algebra. 
Such metric conserving transformations are of physical importance, since 
they should describe various physically meaningful 
 unitary transformations on fuzzy spaces including the ones corresponding to the diffeomorphisms in
 usual spacetimes \cite{Sasakura:2009hs}.
It is not generally possible to obtain such metric conserving transformations in terms of products of functions as
in the 3-ary operation of this paper, if the algebra of functions does not satisfy the cyclic property. 
Moreover, the cyclic property makes it possible to systematically generate metric conserving transformations
in terms of n-ary \cite{Sasakura:2011nj} and supersymmetric \cite{Sasakura:2011qg} operations.    
It should be interesting to systematically study various aspects of spacetime symmetry transformations 
on physically motivated fuzzy spacetimes from this new perspective.
It would also be possible that the algebraic structure presented in this paper may have some applications 
to the description of the nonassociative fuzzy spacetimes discussed in D-brane 
setups \cite{deMedeiros:2004wb,Ramgoolam:2003cs,Ramgoolam:2001zx}.
 
While commutation relations of coordinates can well describe some symmetric noncommutative spacetimes
such as the noncommutative two-sphere \cite{Madore:1991bw}, n-ary algebras of coordinates can be expected to
characterize symmetric fuzzy spaces with nonassociativity 
\cite{deMedeiros:2004wb,Ramgoolam:2003cs,Ramgoolam:2001zx}.
To pursue this expectation further, the commutative nonassociative fuzzy flat spacetimes
motivated from the tensor models have been investigated, 
and a 3-ary relation of coordinates has been found for the fuzzy flat spacetimes.
While the Lorentz symmetry is obviously satisfied in the obtained 3-ary relation, 
it apparently contradicts with the translational symmetry of the fuzzy
flat spacetimes. But a detailed analysis of the flat spacetime algebra has shown that
the 3-ary relation is indeed consistent with a 3-ary translational transformation.
This aspect is in good parallel with the fact that the apparently violated translational symmetry of a 
three-dimensional noncommutative space obtained from lattice gravity
is indeed conserved by a Hopf algebraic translation \cite{Freidel:2005me,Freidel:2005bb,Sasai:2007me}. 

The 3-ary relation of the coordinates of the fuzzy flat spacetimes has been shown to form a Lie triple system,
and the associated Lie algebra has been shown to be identical with the Snyder's noncommutative spacetime.
A similar result has been obtained in another kind of nonassociative
spacetimes in Refs.~\onlinecite{Girelli:2010zw,Girelli:2010wi}.
These examples show interesting transmutation from nonassociativity to noncommutativity, and,
as stressed in Refs.~\onlinecite{Girelli:2010zw,Girelli:2010wi}, can provide a solution to
the problem of unwanted dimensions in noncommutative spacetimes.
It would be interesting to study whether other noncommutative spacetimes can also be embedded into
nonassociative ones in similar manners.

What seems especially interesting is the appearance of the 3-ary product, by which the symmetry of the rank-three
tensor models can be represented. 
This new way of describing symmetries by 3-ary algebras seems to contain a conceptual interest in 
origins of symmetries. In a usual framework, a symmetry is a given input rather than an emergent phenomenon.
On the other hand, however, the form of the symmetry transformation by the 3-ary algebra in the tensor models
 suggests that
the symmetry can only appear when a background is generated. This new aspect of symmetry should be 
explored further in the context of the general idea that spacetime and gravity are emergent phenomena
\cite{Verlinde:2010hp,Padmanabhan:2008zza,Berenstein:2005aa,Horowitz:2006ct,Lin:2004nb,
Lee:2010zf,Steinacker:2010rh,Berenstein:2008eg,Erdmenger:2007xs,
Seiberg:2006wf,Kawai:2002jk,Oriti:2006ar}.

\begin{acknowledgments}
Part of the contents of this paper has first been presented in the lecture at 
``Shinshu Winter School'', held March 10-12, 2011, in
Shiga Heights Villa, Ochanomizu University.
The author would like to thank the organizers of this winter school for 
giving him the opportunity to get the main ideas of this paper on this occasion.
The author would also like to thank M. Axenides,  E. Floratos,  F. Girelli, and S. Ramgoolam for 
e-mail correspondence.  
\end{acknowledgments}

\bibliography{aip}% Produces the bibliography via BibTeX.

\end{document}